\documentclass[twocolumn,amsmath,amssymb,prl]{revtex4}
\usepackage{graphicx}

\newcommand{\ra}{\rangle}

\newcommand{\ket}[1]{\left|#1\right\rangle}

\begin{document}

\title{Environment Assisted Precision Measurement}

\author{G. Goldstein$^{1}$, P. Cappellaro$^{1,2,3}$, J.R. Maze$^{1}$,
J. S. Hodges$^{1,3}$, L. Jiang$^{1,5}$, A. S. S{\o}rensen$^{4}$,
M. D. Lukin$^{1,2}$}

\affiliation{$^{1}$Department of Physics, Harvard University, $^{2}$Harvard-Smithsonian
Center for Astrophysics, Cambridge MA 02138 USA}

\affiliation{$^{3}$Nuclear Science and Engineering Dept., Massachusetts Institute
of Technology, Cambridge MA 02139 USA}

\affiliation{$^{4}$ QUANTOP, Niels Bohr Institute, Copenhagen University, DK
2100, Denmark} 
\affiliation{$^{5}$Institute of Quantum Information, California
Institute of Technology, Pasadena CA 91125 USA }

\begin{abstract}
We describe a method to enhance the sensitivity of precision measurements
that takes advantage of a quantum sensor's environment to amplify
its response to weak external perturbations. An individual qubit is
used to sense the dynamics of surrounding ancillary qubits, which
are in turn affected by the external field to be measured. The resulting
sensitivity enhancement is determined by the number of ancillas that
are coupled strongly to the sensor qubit; it does not depend on the
exact values of the coupling strengths and is resilient to many forms
of decoherence. The method achieves nearly Heisenberg-limited precision
measurement, using a novel class of entangled states. We discuss specific
applications to improve clock sensitivity using trapped ions and magnetic
sensing based on electronic spins in diamond. 
\end{abstract}
\maketitle

Precision measurement is among the most important applications of
resonance methods in physics. For example, quantum control of atomic
systems forms the physical basis of the world's best clocks. Ideas
from quantum information science have been used to demonstrate that
quantum entanglement can enhance these measurements \cite{key-62,key-63}.
At the same time a wide range of quantum systems have been recently
developed aimed at novel realizations of solid state qubits. Potentially
such systems can be used as quantum measurement devices such as magnetic
sensors with a unique combination of sensitivity and spatial resolution
\cite{key-35,key-36,key-41,Bouchard09}. In this Letter, we describe
a novel technique that makes use of the sensor's local environment
as a resource to amplify its response to weak perturbations. We shall
use solid state sensors and ion clocks as examples.

The purpose of quantum metrology is to detect a small external field,
coupled to the sensor by a Hamiltonian: $H_{b}^{S}=b(t)\kappa S_{z}$,
where $S_{z}$ is the spin operator of the quantum sensor. Here, $b(t)$
can be an external magnetic field or the detuning of a laser from
a clock transition while $\kappa$ is the spin's coupling to the field.
The working principle of almost any quantum metrology scheme can be
reduced to a Ramsey experiment \cite{key-47,key-48}, where the field
is measured via the induced phase difference between two states of
the quantum sensor. The figure of merit for quantum sensitivity is
the smallest field $\delta b_{min}$ that can be read out during a
total time $T$. For a single spin 1/2, if the sensing time is limited
to $\tau$ (e.g. by environmental decoherence) then: $\delta b_{min}\simeq\frac{1}{\kappa\sqrt{T\tau}}$. 

In many cases the external field also acts on the sensor's environment,
which normally only induces decoherence and limits the sensitivity.
Here we show that in some cases the environment can instead be used
to enhance the sensitivity. For generality we will illustrate the
key ideas using the \textit{central spin model} (Fig. 1a). In this
model a central spin (which can be prepared in a well defined initial
state, coherently manipulated and read out) is coupled to a bath of
\textit{dark} spins that can be polarized and collectively controlled,
but cannot be directly detected. The system is described by the Hamiltonian
$H=H_{b}+H_{int}$, with \begin{equation}
H_{b}=b\left(t\right)\left(\kappa S_{z}+\xi\sum I_{z}^{i}\right),\ \ H_{int}=|1\rangle\langle1|\sum\lambda_{i}I_{z}^{i},\label{eq:Main Hamiltonian}\end{equation}
 where $\lambda_{i}$ are the couplings between sensor and environment
spins, while $\kappa$ and $\xi$ are couplings to the external field
of the central and dark spins respectively. Here $\ket{0}$, $\ket{1}$
and $S_{z}$ refer to the central spin while $\ket{\uparrow}$, $\ket{\downarrow}$,
$I_{z}^{i}$ to the dark spins (and we set $\hbar=1$). We consider
two cases. In the first one $H_{int}$ can be turned on and off at
will and is much larger than any other interaction in the system (e.g.
a laser-mediated ion interaction). In the second case, $H_{int}$
is intrinsic to the system and of the same order of magnitude as the
relevant sensing time (e.g. dipole-dipole interactions between solid
state spins). In all cases we will assume collective control over
the dark spins.

To illustrate the sensing method we consider first the idealized case
where the couplings between the central and the dark spins can be
turned on and off at will and the dark spins are initialized in a
pure state $\left|\uparrow\uparrow...\uparrow\right\rangle $. 
\begin{figure}[hb]
\centering 
\includegraphics[scale=0.8]{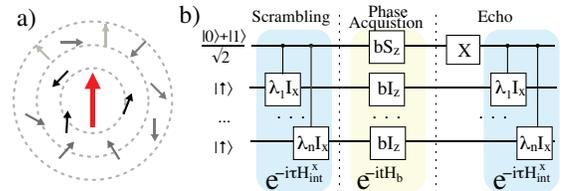} 
\caption{\label{fig:CentralSpin} \textbf{Ideal Model.} (a) A central spin
coupled to a spin bath. (b) Ideal measurement procedure. Gates labeled
$\lambda I_{x}$ represent controlled rotations $e^{-i\lambda I_{x}}$
of the dark spins, obtained via the Hamiltonian $H_{int}$. The gates
$bI_{z}$ represent rotations $e^{-ibI_{z}}$ due to the external
field $b$. The central spin undergoes a spin-echo before measurement.}
\end{figure}

Consider the circuit in Fig. 1(b). First, the central spin is prepared
in an equal superposition of the two internal states $\sim|0\ra+|1\ra$.
Then $H_{int}$ is rotated to the $x$-axis and applied for a time
$\tau$. 
 This induces controlled rotations of the dark spins, resulting in
an entangled state

\begin{equation}
\left(|0\rangle|\uparrow...\uparrow\rangle+|1\rangle|\varphi_{1}...\varphi_{N}\rangle\right)/\sqrt{2},\label{eq:state}\end{equation}
 where $|\varphi_{i}\rangle\equiv\cos\left(\varphi_{i}\right)\ket{\uparrow}-i\,\sin\left(\varphi_{i}\right)\ket{\downarrow}$
with $\varphi_{i}=\lambda_{i}\tau$. This state is then used to sense
the magnetic field. We assume for now $\kappa=0$ and define $\theta_{d}=\xi\,\int_{0}^{\tau}dt\, b(t)$.
Under the action of the magnetic field the state evolves to $\left(\ket{0}\ket{\uparrow\dots}+\ket{1}\ket{\psi_{1}\dots\psi_{N}}\right)/\sqrt{2}$
with 
\[
\begin{array}{ll}
\ket{\psi_{i}}= & \cos\varphi_{i}\ket{\uparrow}_{i}-i\sin\varphi_{i}e^{2i\theta_{d}}\ket{\downarrow}_{i}\\
 & \approx e^{i2\theta_{d}sin^{2}\varphi_{i}}\ket{\varphi_{i}}+\theta_{d}\frac{\sin(2\varphi_{i})}{2}\ket{\varphi_{i}^{\perp}},
 \end{array}
 \]
 where $\left\langle \varphi_{i}|\varphi_{i}^{\perp}\right\rangle =0$.
The central spin is then flipped by a $\pi$-pulse and another control
operation with $H_{int}$ along $x$ is applied. If the field (and
thus $\theta_{d}$) were zero the interaction between central and
dark spins would then be refocused, corresponding to decoupling the
sensor spin from the environment as in a spin-echo. For metrology
purposes, to first order, the effect of a small field is to introduce
a phase difference $\Phi\approx2\theta_{d}\sum_{i}\sin^{2}(\varphi_{i})$
between the states of the dark spins, depending on the state of the
central spin (while the terms $\propto\ket{\varphi_{i}^{\perp}}$
only contribute to second order in $\theta_{d}$). After the sensor
spin is rotated around the $x$-axis, this yields an additional contribution
to the probability of finding the spin in the $\ket{1}$-state: $P_{1}\approx(1+\Phi+\theta_{s})/2+O(b^{2})$
(where we reintroduced the phase acquired by the sensor spin alone,
$\theta_{s}=\kappa\int_{0}^{\tau}dt\, b(t)$, which can be simply
added as $H_{b}^{S}$ commutes with the rest of the Hamiltonian).

While the signal is enhanced by a factor $\propto\Phi$, the quantum
projection noise remains the same as we still read out one spin only.
The minimum field that can be measured in a total time $T$ is then:
\begin{equation}
\delta b_{min}=\sqrt{\frac{\tau}{T}}\frac{1}{\Phi+\theta_{s}}\approx\frac{1}{n\xi\sqrt{T\tau}},
\label{eq:SensitivityCNOT}
\end{equation}
 where $n$ is the total number of dark spins. The linear scaling
in $n$ of the phase $\Phi$ can be achieved in principle for any
distribution of $\lambda_{i}$'s, since we can always choose a duration
$\tau$ such that $\left\langle \sin^{2}\left(\lambda_{i}\tau\right)\right\rangle \geq\frac{1}{2}$,
leading to order one contribution from each spin. Thus we are able
to perform Heisenberg-limited spectroscopy despite the fact that the
precise form of the entangled state (\ref{eq:state}) is uncontrolled,
and may not even be known to us~\cite{Cappellaro05}. This considerably
relaxes the requirements for entanglement enhanced spectroscopy as
compared to known strategies involving squeezed or GHZ-like states.
We next discuss two experimental implementations that approximate
this idealized scheme: quantum clocks with trapped ions and spin-based
magnetometry.

To reach high precision in quantum clocks, the ions must posses several
characteristics: a stable clock transition, a cooling cycling transition,
good initial state preparation and reliable state detection. It is
then convenient to use two species of ions in the same Paul trap~\cite{key-5,key-59}:
The \textit{spectroscopy} ions (e.g. $^{27}Al^{+}$) provide the clock
transition while the \textit{logic} ion (e.g. $^{9}Be^{+}$) fulfills
the other requirements. Although inspired by a similar idea as the
one proposed here, experiments using two ion species have been so
far limited to just one {spectroscopy} ion~ \cite{key-59,key-4,key-3,key-5};
with our method the number of spectroscopy (\textit{dark}) ions can
be increased.

Specifically, by using multichromatic gates~\cite{Sorensen} one
can implement the Hamiltonian $H_{int}$ in Eq. (\ref{eq:Main Hamiltonian})
\cite{key-64} (such multichromatic gates are known to be much more
robust to heating noise than Cirac-Zoller gates~\cite{Sorensen}
that have been used so far). We can then use the method presented
here to transfer the phase difference due to the detuning of several
spectroscopy ions onto a single logic ion, which can be read out by
fluorescence. In principle, we achieve Heisenberg-limited sensing
of the clock transition without individual addressability of the spectroscopy
ions or producing GHZ states. Importantly, we achieve this even if
the spectroscopy ions have different couplings to the logic ion, which
will be the case in a trap with different ion species due to the absence
of a common center of mass mode.

We next briefly discuss the effects of decoherence on this method.
It has been argued (see e.g. \cite{key-1}), that the coherence time
reduction for a $n$-spin entangled state reduces the sensitivity
to roughly that of spectroscopy performed on $n$ individual spin,
scaling as $\sqrt{n}$, as opposed to the ideal scaling $\propto n$
derived here (See Ref. \cite{key-7} for a different scaling in the
case of atomic clocks). Our scheme would obtain an improvement $\propto\sqrt{n}$
with respect to current experimental realizations where only a single
ancillary ion is available, even in the decoherence model of Ref.
\cite{key-1}, where each spin undergoes individual Markovian dephasing.
Furthermore, this decoherence model is not so relevant in present
setups, as technical noise during the gates and imperfect rotations
are dominant for traps with many ions. Our method is highly robust
to static repetitive imperfections during the gates, leading to further
improvement depending on the exact noise specifications \cite{key-64}.

In many physical situations short bursts of controlled rotations,
as used above, are not available. Instead, the couplings between the
central and dark spins are always on and their exact strength is unknown.
Examples of such systems are solid-state spin systems used for magnetometry~\cite{key-35,key-36,key-41,Bouchard09}.
Still, it is possible to achieve nearly Heisenberg-limited metrology
even for these systems. 

Specifically, we will consider magnetic sensing using a single Nitrogen
Vacancy (NV) center in diamond~\cite{key-37}, surrounded by \textit{dark}
spins associated with Nitrogen electronic impurities~\cite{key-39,key-40}.
We focus on NV centers since their electronic spins (S=1) can be efficiently
initialized into the $S_{z}=0$ state by optical pumping and measured
via state selective fluorescence. By applying an external (static)
magnetic field that splits the degeneracy between $S_{z}=\pm1$ states
and working on resonance with the $(0\leftrightarrow1)$ transition,
the NV center can be reduced to an effective two-level system~\cite{key-36}.
Then, the system comprising one NV center and several N spins is well
described by Hamiltonian (\ref{eq:Main Hamiltonian}).

\begin{figure}[htb]
\centering 
\includegraphics{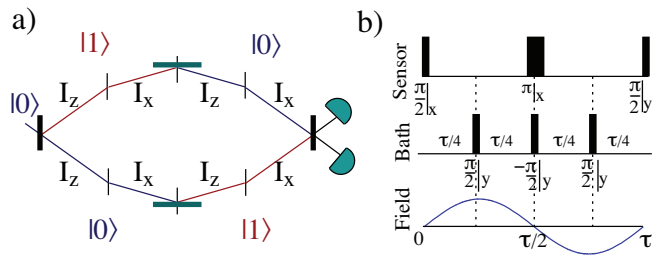} 
\caption{\label{fig:Sequences}\textbf{Method.} a) Mach-Zehnder interferometer,
showing the central spin state and the effective Hamiltonian of the
dark spins along each arm. b) Pulse sequence for environment assisted
magnetometry.}
\end{figure}

In Fig. \ref{fig:Sequences} we introduce a control sequence that
yields an effect equivalent to the circuit in Fig. 1.b. The action
of the pulse sequence can be best understood using the well known
equivalence between Ramsey spectroscopy and Mach-Zehnder interferometry
\cite{key-47,key-48}, where the interferometer arms describe the
central spin state. It is sufficient to consider the evolution of
each arm separately, replacing $S_{z}$ by its eigenvalues $m_{s}=\left\{ 0,\,1\right\} $
and describing the evolution in the interaction frame defined by the
control pulses~\cite{key-46}. Hamiltonian (1) becomes time-dependent,
with dark spins alternating between $I_{z}^{i}$ and $I_{x}^{i}$
as shown in Fig. \ref{fig:Sequences}. Then, for different halves
of the spin echo sequence, the coupling Hamiltonian in each arm is
zero ($m_{s}=0$) while for the other halves it has identical forms.
In the absence of a magnetic field the evolution is thus the same
along each arm of the interferometer. Adding an external field creates
a phase shift between the two arms. For small field strengths we can
then evaluate the phase difference acquired between the two arms as
a perturbation. 
For a finite polarization $P$ of the dark spins we find \begin{equation}
\Phi=\overline{\theta_{s}}\left[1+2P\sum\frac{\overline{\theta_{d}}}{\overline{\theta_{s}}}\sin^{2}(\varphi_{i}/4)\right],\label{eq:phase}\end{equation}
 with $\overline{\theta_{s}}=\frac{\kappa}{\tau}\left[\int_{0}^{\frac{\tau}{2}}dt\, b(t)-\int_{\frac{\tau}{2}}^{\tau}dt\, b(t)\right]$,
$\overline{\theta_{d}}=\frac{\xi}{\tau}\int_{\frac{\tau}{2}}^{\frac{3\tau}{4}}dt\, b\left(t\right)$.
Compared to spin-echo based magnetometry~\cite{key-35}, the signal
is increased by the factor in the square bracket while the measurement
noise is the same, as we still read out one spin only. Note that all
the dark spins contribute positively. For values of the couplings
such that $\left|\lambda_{i}\tau\right|\geq\pi$, or \textit{strongly
coupled} environment spins, we obtain a contribution $\propto~2n_{sc}\,\langle\sin^{2}(\frac{\lambda_{i}\tau}{4})\rangle\approx n_{sc}$.
Each of the $n_{sc}$ strongly coupled spins thus gives a contribution
of order one, irrespective of the sign or exact value of the coupling.
\textit{Weakly coupled} dark spins ($\lambda_{i}\tau\leq1$) contribute
instead with a factor $(\lambda_{i}\tau)^{2}/8$ and we obtain a total
phase $\Phi\approx\overline{\theta_{s}}\left[1+\frac{\overline{\theta_{d}}}{\overline{\theta_{s}}}P\left(n_{sc}+\frac{1}{8}\sum^{'}\left(\lambda_{i}\tau\right)^{2}\right)\right]$,
where the primed sum is on the weakly coupled spins. In general the
sensitivity enhancement scales as $\sim P\, n_{sc}$. We thus achieve
nearly Heisenberg-limited sensing of the external field.

We next take into account decoherence resulting from the interaction
with the environment spins as well as decoherence of the dark spins
themselves. To evaluate these effects, we compare the sensitivity
achievable with the proposed pulse sequence to that obtained with
a spin-echo sequence, considering the same system (a sensor spin surrounded
by the same spin bath) and including the effects of decoherence (external
perturbations and couplings between the dark spins). Once the central
spin looses phase coherence due to interactions with the bath, it
is no longer possible to use it for magnetometry. This limits the
sensing time and consequently the magnetometer sensitivity. Spin echo
(as well as more sophisticated decoupling techniques \cite{key-46,key-45,key-5,key-64})
can be used to prolong the phase coherence of the central spin. Under
realistic assumptions the coherence time for the pulse sequence presented
in this Letter is on the same order of the sensor coherence time under
spin-echo $T_{2}^{s}$. In essence, the decoherence rate of an entangled
state of the form (\ref{eq:state}) is dictated by the internal evolution
of the strongly coupled spins, and the relevant decoherence time is
thus the time it takes before any of these spins have decohered. The
same is, however, the case for spin echo sequences: if a central spin
is strongly coupled to $n_{sc}$ dark spins, a spin flip of a single
dark spin will lead to different evolutions in the two halves of the
spin echo sequence, thus decohering the state of the central spin.
As the source of both decoherence processes is the dipole-dipole interaction
among dark spins, the coherence times of spin-echo and our procedure
are on the same order and the signal amplification obtained with our
strategy (Eq. \ref{eq:phase}) thus yields a sensitivity enhancement~\cite{footnote2}.

\begin{figure}[t]
\centering
\includegraphics[scale=0.18]{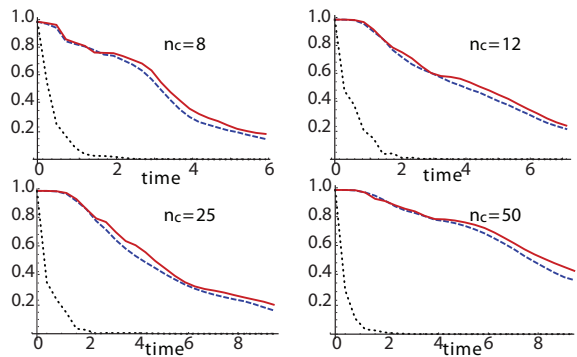} 
\caption{\label{fig:SignalDecay}\textbf{Simulations.} Normalized signal decay
for the proposed (red, solid lines), spin-echo sequence (blue, dashed
lines) and no control (black, dotted lines). We assumed perfect delta
function pulses. A leading order cluster expansion was used~\cite{key-51}.
20 dark spins were randomly placed in a cube of side-length $\sqrt[3]{20}$
with the sensor spin at the center. We set $g\mu_{B}\equiv1\left[m\right]^{3/2}\left[s\right]^{-1/2}$
to use dimensionless quantities. WAHUHA sequences~\cite{key-46}
with $n_{c}$=8, 12, 25 and 50 cycles per echo interval were simulated.
Each curve is an average over 10 Monte Carlo simulations. For simplicity
we set $P=0$.}
\vspace{-10pt}
\end{figure}

To be more quantitative, we analyze the evolution due to dipole-dipole
couplings in the bath, described by the Hamiltonian \begin{equation}
H=|1\rangle\langle1|\sum\lambda_{i}I_{z}^{i}+\sum_{i<j}\kappa_{ij}\left(3I_{z}^{i}I_{z}^{j}-\overrightarrow{I_{i}}\cdot\overrightarrow{I_{j}}\right)\label{eq:DecoherenceHamiltonian}\end{equation}
 A short time expansion shows that the state fidelity at the end of
the pulse sequence is given by $1-{\displaystyle \frac{\tau^{4}}{8}\cdot\sum_{i<j}}\left(1-P^{2}\right)\kappa_{ij}^{2}\left(\lambda_{i}-\lambda_{j}\right)^{2}+...$,
and $1-\frac{\tau^{4}}{128}{\displaystyle \sum_{i<j}}\kappa_{ij}^{2}\times\left(7(\lambda_{i}^{2}+\lambda_{j}^{2})+12\lambda_{i}\lambda_{j}+P^{2}\left(\lambda_{i}+2\lambda_{j}\right)\left(2\lambda_{i}+\lambda_{j}\right)\right)+...$,
for the spin-echo and modified pulse sequences, respectively. For
experimentally relevant values of polarizations, similar reductions
in fidelity occur for both pulse sequences \cite{key-8}. To go beyond
the short time expansions we have simulated the signal decay for both
spin-echo and the proposed pulse sequence (see Fig. \ref{fig:SignalDecay}).
We compared these results to the signal decay when no control sequence
is applied (for the same environment the decay is now described by
the dephasing time $T_{2}^{*}$). We simulated a spin bath composed
of spin 1/2 paramagnetic impurities, undergoing a WAHUHA sequence
(which is designed to refocus the dipole-dipole coupling of the environment
spins, but does not cancel out the coupling to the external field~\cite{key-46,key-64}).

A different limitation on the sensing time $\tau$ is set by the fact
that the orientation of the dark spin will not be static as assumed.
Dipole-dipole couplings among dark spins during each $\tau/4$ period
of free evolution rotate each spin away from the initial direction.
This rotation means that the spins cease to build up a phase difference
between the two arms of the interferometer for time scales comparable
to the correlation time of the dark spin bath $\tau_{c}^{d}$. 
The optimum sensing time is thus $T\sim\min\left\{ T_{2}^{s},\,\tau_{c}^{d}\right\} $.
Since in most systems $\tau_{c}^{d}\geq T_{2}^{s}$ \cite{key-46},
the optimum sensing time of this pulse sequence is comparable to that
of spin-echo based magnetometry, thus the sensitivity enhancement
is roughly the same as the signal strength enhancement. 

In conclusion, we proposed a scheme to enhance precision measurement
by exploiting the possibility to coherently control the ancillary
qubits. In solid state implementations we are able to exploit dark
spins in the bath while preserving roughly the same coherence times
as in spin-echo based magnetometry. Thus signal enhancement leads
directly to sensitivity enhancement. For trapped ion implementations
we can use imperfect phase gates and still achieve Heisenberg-limited
sensitivity. Our method has the potential to be applied more generally,
using different systems and more sophisticated pulse sequences \cite{key-64}.
It opens the possibility to use a broad class of partially entangled
states to achieve Heisenberg limited metrology, even in the presence
of disordered couplings, partial control and decoherence. \\
 \textbf{Acknowledgments}. This work was supported by the NSF,
ITAMP, DARPA and the Packard Foundation.


\begin{thebibliography}{35}
\bibitem{key-62}P. O. Schmidt, \textit{et al}., Science \textbf{309},
749 (2005)

\bibitem{key-63}M. H. Schleier-Smith \textit{et al}., Arxiv 0810.2583

\bibitem{key-35}J. M. Taylor, \textit{et al}., Nat. Phys. \textbf{4},
810 (2008)

\bibitem{key-36}J. R. Maze, \textit{et al}., Nature \textbf{455},
644 (2008)

\bibitem{key-49}G. Balasubramanian, \textit{et al}., Nature 445,
\textbf{644} (2008)

\bibitem{key-53}P. Neumeann, \textit{et al}., Science \textbf{320},
1326 (2008)

\bibitem{key-54}V. Jacques, \textit{et al}., Phys. Rev. Lett. \textbf{102},
0547403 (2009)

\bibitem{key-37}R. J. Epstein, \textit{et al}., Nat. Phys. \textbf{1},
94 (2005)

\bibitem{key-38}M. V. G. Dutt, \textit{et al}., Science \textbf{316},
1312 (2007)

\bibitem{key-39}R. Hanson, \textit{et al}., Phys. Rev. Lett. \textbf{97},
087601 (2006)

\bibitem{key-40}T. Gaebel, \textit{et al}., Nat. Phys. \textbf{2},
408 (2006)

\bibitem{key-41}S. L. Braunstein, \textit{et al}., Ann. Phys. (San
Diego) \textbf{247}, 135 (1996)

\bibitem{key-42}S. M. Roy and S. L. Braunstein, Phys. Rev. Lett.
\textbf{100}, 220501 (2008)

\bibitem{key-43}J. J. Bollinger, \textit{et al}., Phys. Rev. A \textbf{54},
R4649 (1996)

\bibitem{key-44}K. Khodjasteh and D. A. Lidar, Phys. Rev. A \textbf{75},
062310 (2007)

\bibitem{key-45}P. Mansfield Journal Physics C \textbf{4}, 1444 (1971)

\bibitem{key-46}U. Haeberlen, \textit{High Resolution NMR in Solids
Selective Averaging} (Academic Press Inc., 1976)

\bibitem{key-47}D. J. Wineland, \textit{et al}., Phys. Rev. A. \textbf{50},
67 (1994)

\bibitem{key-48}B. Yurke, \textit{et al}., Phys. Rev. A. \textbf{33},
4033 (1986)

\bibitem{Cappellaro05}P. Cappellaro, \textit{et al}., Phys. Rev.
Lett. \textbf{94}, 020502 (2005)

\bibitem{Sorensen} A. S{\o}rensen, K. M{\o}lmer, Phys. Rev.
A. \textbf{62}, 022311 (2000)

\bibitem{key-52}W.A. Coish and D. Loss, Phys. Rev. B \textbf{72},
125337 (2005)

\bibitem{key-51}W. M. Witzel and S. D. Sarma, Phys. Rev. B \textbf{74},
035322 (2006)

\bibitem{key-59}T. Rosenband \textit{et al}., Phys. Rev. Lett. \textbf{98},
220801 (2007)

\bibitem{key-61}S. Boixo and R. D. Somma, Phys. Rev. A \textbf{77},
052320 (2008)

\bibitem{key-64}G. Goldstein \textit{et al}., In preparation. 

\bibitem{key-2}M. Kitagawa and M. Ueda, Phys Rev A \textbf{47}, 5138
(1993)

\bibitem{key-3}D. Liebfried \textit{et al}., Science \textbf{304},
1476 (2004)

\bibitem{key-4}D. Liebfried \textit{et al}., Nature \textbf{438},
638 (2005)

\bibitem{key-5}D. Liebfried \textit{et al}., Nature \textbf{422},
412 (2003)

\bibitem{key-6} Note that the present method scales favorably with
dark spins polarization ($P$) As seen, the sensitivity scales as
$\left(P\, N\right)^{-1}$ while it would scale as $(P^{N}N)^{-1}$
for a GHZ state and as $\frac{1-P^{2}}{\sqrt{N}}$ for spin squeezed
states.

\bibitem[31]{key-1}S. F. Huelga \textit{et al}., Phys Rev Lett \textbf{79},
3865 (1997)

\bibitem[32]{key-7}A. Andre \textit{et al.}, Phys Rev Lett \textbf{92},
230801 (2004) 

\bibitem[33]{Bouchard09} L-S. Bouchard, V. M. Acosta, E. Bauch, D.
Budker, Arxiv:0911.2533 (2009) 

\bibitem[34]{footnote2} Although our pulse sequence is more sensitive
to dark spin dephasing (since it creates superposition states of the
dark spins), this noise source is not relevant in current experiments,
where noise is dominated by single species dipole-dipole interactions~\cite{key-3,key-4,key-5}. 

\bibitem[35]{key-8}For $P\approx1$ spin-echo achieves longer coherence
times than the proposed sequence, since flip-flops are quenched in
a fully polarized bath, they would still be allowed in the proposed
pulse sequence. However the improvement in coherence times for spin
echo scales poorly as $\sim\sqrt[4]{1-P^{2}}$ and homonuclear decoupling
sequences such as WAHUHA can be used to reduce the flip-flop effects.
\end{thebibliography}
\end{document}